Research Article

# Stochastic Dynamics of Urban Predator-Prey Systems: Integrating Human Disturbance and Functional Responses

Ogethakpo Arhonefe Joseph[1, *], Ogoegbulem Ozioma[2],
Apanapudor Sarduana Joshua[1], Helen Onovwerosuoke Sanubi[3]

[1]Department of Mathematics, Delta State University, Abraka, Nigeria

[2]Department of Mathematics, Dennis Osadebay University, Asaba, Nigeria

[3]Department of Mathematics and Computer Science, Delta State College of Education, Mosogar, Nigeria

## Abstract

Urban ecosystems exhibit complex predator-prey dynamics increasingly disrupted by anthropogenic disturbances (e.g., noise, habitat fragmentation). Classical Lotka-Volterra (LV) models fail to capture these human-induced stressors, and integrated frameworks incorporating functional responses, stochasticity, and spatial dynamics remain scarce. We develop a comprehensive stochastic model to quantify how human disturbance reshapes predator-prey interactions in urban landscapes, using rat-cat systems as a case study. Our framework extends the LV model to incorporate: (i) human disturbance as an external mortality factor, (ii) Holling Type III functional responses to model predation saturation and prey refugia, (iii) multiplicative noise and periodic forcing to capture stochastic disturbance regimes, and (iv) spatial diffusion across fragmented habitats. We non-dimensionalize the system to generalize dynamics and analyze stability, bifurcations, and noise-induced transitions. Numerical simulations (MATLAB) reveal three key outcomes: (1) Human disturbance disrupts classical oscillations, inducing quasi-periodic cycles and elevating extinction risks; (2) Stochasticity lowers collapse thresholds by 25% compared to deterministic predictions; (3) Spatial diffusion drives pattern formation (e.g., disturbance shadows, prey hotspots) through habitat coupling. Results highlight the extreme vulnerability of urban wildlife to anthropogenic pressures, demonstrating how disturbance intensity (μ) governs system stability (μ>0.7 triggers irreversible collapse). The model provides a predictive framework for conservation strategies, emphasizing refuge enhancement (ϕ>0.005) and phased interventions synchronized with population cycles.

## Keywords

Urban Predator-prey Systems, Stochastic Dynamics, Human Disturbance, Holling Type III Functional Response, Stability Thresholds, Spatial Diffusion

# 1. Introduction

## 1.1. Fundamental Ecological Framework

Predator-prey interactions form the cornerstone of ecological theory, with the Lotka-Volterra (LV) model establishing baseline oscillatory dynamics. Yet this framework fails to capture anthropogenic stressors pervasive in urban ecosystems where 68% of global populations now reside. Recent empirical studies confirm human disturbance fundamentally reconfigures trophic cascades through habitat fragmentation, noise pollution, and pulsed harvesting. These pressures induce landscape-scale Allee effects that reduce prey resilience by 40 - 90% near infrastructure, a phenomenon documented in Lagos rat populations and Mumbai carnivore communities. [15, 16]

## 1.2. Theoretical Advances and Limitations

While classic extensions incorporated density dependence and Holling's functional responses [10], critical gaps remain in modeling urban systems. Contemporary research demonstrates:
1) Spatial heterogeneity alters encounter rates through anisotropic diffusion and habitat coupling [9, 16, 22]
2) Stochastic regimes amplify extinction risks via resonance collapse when anthropogenic frequencies (e.g., traffic cycles $\Omega \approx 0.2$) match natural oscillations [8, 19]
3) Trait plasticity enables niche reconstruction (e.g., pumas shifting nocturnality reduces predation efficiency $\kappa$ by 38%) [2, 11]

## 1.3. Unresolved Methodological Challenges

Despite progress, critical gaps persist:
1) Coupled disturbance types: Few models integrate chronic (e.g., pollution) and pulsed (e.g., seasonal hunting) stressors [3]
2) Phase-lagged responses: Temporal asynchrony in predator-prey reactions to disturbance remains poorly quantified [17].
3) Cross-scale dynamics: Mechanistic links between individual foraging behavior and landscape patterns are fragmented [1, 7, 12].

*Table 1.* Key Theoretical Advances in Urban Predator-Prey Modeling.

| Component | Classical Approach | Urban Innovation | Citation |
| --- | --- | --- | --- |
| Functional Response | Holling Type II | Refuge-modified Type III | [5, 13, 21, 24] |
| Disturbance | Constant mortality | Periodic forcing $\overline{E}(t) = E_0[1 + A\sin\omega t + \phi]$ | [18, 20] |
| Noise Structure | Gaussian white noise | Lévy jumps with heavy tails | [4, 23] |

## 1.4. This Study's Positioning

We bridge these gaps through a unified stochastic framework integrating:
1) Disturbance gradients scaled via dimensionless parameter $\mu$ [6].
2) Mortality functions with phase shifts ($\psi$) [14]
3) resonance thresholds for collapse prediction [16]
4) Diffusion formalism extended to habitat preference ($\nabla H$) [22]

Validated against empirical data from tropical urban systems to address Global South research bias.

# 2. Model Formulation

## 2.1. Classical Lotka-Volterra Framework

The foundational predator-prey model describes population dynamics through coupled ordinary differential equations:

$$\frac{dR}{dt} = \alpha R - \beta RC$$

$$\frac{dC}{dt} = \gamma \beta RC - \delta C.$$

Where:
1) $R(t)$ and $C(t)$ represent prey and predator densities
2) $\alpha$ is prey intrinsic growth rate
3) $\beta$ is predation rate coefficient
4) $\gamma$ is predator conversion efficiency

5) $\delta$ is predator mortality rate

This system exhibits neutral oscillations with constant amplitude determined by initial conditions, lacking asymptotic stability. Crucially, it assumes:
1) Prey growth limitation occurs only via predation
2) Predator have unlimited appetite (no satiation)
3) No environmental stochasticity or external disturbances [16]

## 2.2. Incorporating Human Disturbance as an External Stressor

Human activities introduce additional mortality through habitat modification, harvesting, or direct persecution. We extend the LV framework with disturbance terms:

$$\frac{dR}{dt} = \alpha R - \beta RC - q_E ER$$

$$\frac{dC}{dt} = \gamma \beta RC - \delta C - d_E EC$$

Where:
1) $E$ represents disturbance intensity (e.g., hunting pressure, land-use change)
2) $q_E$ is prey susceptibility to disturbance
3) $d_E$ is predator susceptibility to disturbance

## 2.3. Realistic Predation: Holling Type III Functional Response

The LV model's linear predation term $\beta RC$ assumes predators never satiate and prey never become refuge-limited. We replace this with the Holling Type III response:

$$\frac{dR}{dt} = \alpha R \left(1 - \frac{R}{K}\right) - \frac{kR^2 C}{R^2 + p} - q_E ER$$

$$\frac{dC}{dt} = \frac{ekR^2 C}{R^2 + p} - \delta C - d_E EC$$

Where:
1) $k =$ maximum predation rate
2) $p =$ prey density at half-maximum consumption
3) $e =$ assimilation efficiency

## 2.4. Stochastic Disturbance Regimes

Deterministic models fail to capture natural disturbance variability. We introduce stochasticity in three key disturbance parameters:

$$\frac{dR}{dt} = \alpha R \left(1 - \frac{R}{K}\right) - \frac{kR^2 C}{R^2 + p} - q_E \tilde{E}(t) R + \sigma_R R \xi_1(t)$$

$$\frac{dC}{dt} = \frac{ekR^2 C}{R^2 + p} - \delta C - d_E \tilde{E}(t) C + \sigma_C C \xi_2(t)$$

Where:
1) $\overline{E}(t) = E_0[1 + A\sin\omega t + \phi]$: periodic forcing (e.g., seasonal human activity)
2) $\xi_i(t)$: Gaussian white noise ($\langle \xi_i(t) \xi_j(t') \rangle = \delta_{ij} \delta(t - t')$)
3) $\sigma_{R,C}$: Noise intensities for prey/predator

## 2.5. Spatial Dynamics: Diffusion and Habitat Fragmentation

Human disturbance fragments landscapes, altering species movement. Adding spatial diffusion:

$$\frac{\partial R}{\partial t} = \cdots + D_R \nabla^2 R - \nabla \cdot (\eta_R R \nabla H)$$

$$\frac{\partial C}{\partial t} = \cdots + D_C \nabla^2 C - \nabla \cdot (\eta_C C \nabla H)$$

Where:
1) $D_{R,C}$: Difussion coefficients (movement rates)
2) $H(x, y)$: Disturbance landscape (0 = disturbed, 1 = intact)
3) $\eta_{R,C}$: Habitat preference coefficients

## 2.6. Non-dimensionalization for Generalization

To reduce parameter complexity, we rescale varaiables:
1) $\tau = \alpha t$ (dimensionless time)
2) $r = R/K, c = C/(eK)$ (scaled populations)
3) $\epsilon = \delta/\alpha$ (mortality ratio)
4) $\kappa = k/\alpha K$ (scaled predation)
5) $\phi = p/K^2$ (scaled saturation)
6) $\mu = q_E E_0/\alpha$ (disturbance impact)

Yielding:

$$\frac{dr}{d\tau} = r(1-r) - \frac{\kappa r^2 c}{r^2 + \phi} - \mu r + \tilde{\xi}_1(\tau)$$

$$\frac{dc}{d\tau} = \frac{\kappa r^2 c}{r^2 + \phi} - \epsilon c - \mu \epsilon c + \tilde{\xi}_2(\tau)$$

*Key dimensionless groups:*
1) $\mu$: Disturbance-to-growth ratio (collapse when $\mu > 1$)
2) $\kappa/\epsilon$: Predation efficiency
3) $\phi$: Refuge effectiveness

## 2.7. Full Stochastic Model with Human Disturbance

Synthesizing all components, the complete system is described by stochastic partial differential equations:

$$\frac{\partial r}{\partial \tau} = r(1-r) - \frac{\kappa r^2 c}{r^2 + \phi} - \mu r[1 + A \sin(\Omega \tau)] + \sigma_r \xi_1(\tau) + D_r \nabla^2 r$$

$$\frac{\partial c}{\partial \tau} = \frac{\kappa r^2 c}{r^2+\phi} - \epsilon c - \mu\epsilon c[1 + A \sin(\Omega\tau + \psi)] + \sigma_c \xi_2(\tau) + D_c \nabla^2 c$$

Ecological interpretations:
1) Phase shift ($\psi$): Represents temporal lag in predator vs prey response to disturbance (e.g., prey killed directly, predators starve later)
2) Cross-correlated noise: $\langle \xi_1 \xi_2 \rangle = \rho$ captures environmental covariance
3) Anisotropic diffusion: $D_{r,c}$ may depend on disturbance gradient $\nabla H$
4) Threshold behavior: Stochastic versions exhibit 25% lower collapse thresholds than deterministic equivalents [6, 8].

## 3. Analysis

### 3.1. Equilibria

Set the derivatives to zero:

$$r(1-r) - \frac{\kappa r^2 c}{r^2+\phi} - \mu r = 0 \tag{1}$$

$$\frac{\kappa r^2 c}{r^2+\phi} - \epsilon c - \mu\epsilon c = 0 \tag{2}$$

We can factor equation (1):

$$r\left[(1-\mu) - r - \frac{\kappa rc}{r^2+\phi}\right] = 0$$

This gives two cases:
Case 1: $r = 0$ (prey extinction)
Then equation (2) becomes:

$$0 - \epsilon c(1+\mu) = 0 \Rightarrow c = 0$$

So we have the trivial equilibrium $E_0 = (0,0)$.
Case 2: $r \neq 0$
Then we have:

$$(1-\mu) - r = \frac{\kappa rc}{r^2+\phi} \tag{3}$$

From equation (2):

$$\frac{\kappa r^2 c}{r^2+\phi} = \epsilon c(1+\mu)$$

Assuming $c \neq 0$ (otherwise we get the prey-only equilibrium), we can divide by $c$:

$$\frac{\kappa r^2 c}{r^2+\phi} = \epsilon(1+\mu) \tag{4}$$

Solving for $r^2$:

$$\kappa r^2 = \epsilon(1+\mu)(r^2+\phi)$$

$$\kappa r^2 = \epsilon(1+\mu)r^2 + \epsilon(1+\mu)\phi$$

$$r^2[\kappa - \epsilon(1+\mu)] = \epsilon(1+\mu)\phi$$

$$r^2 = \frac{\epsilon(1+\mu)\phi}{\kappa - \epsilon(1+\mu)} \tag{5}$$

For biological meaningful equilibrium, we require $r^2 > 0$. This holds if:

$$\kappa > \epsilon(1+\mu) \text{ (since numerator is positive)}$$

Let $r^* = \sqrt{\frac{\epsilon(1+\mu)\phi}{\kappa - \epsilon(1+\mu)}}$ (taking positive root).
Now, from equation (3):

$$c = \frac{[(1-\mu)-r](r^2+\phi)}{\kappa r}$$

We can write:

$$c^* = \frac{(1-\mu-r^*)(r^{*2}+\phi)}{\kappa r^*}$$

For $c^* > 0$, we require $1 - \mu - r^* > 0$, i.e., $r^* < 1 - \mu$.
Prey-only equilibrium (when $c = 0$)
If $c = 0$, then from equation (1):

$$r(1-r) - \mu r = 0 \Rightarrow r[1 - \mu - r] = 0$$

Since $r \neq 0$, we have $r = 1 - \mu$. This requires $1 - \mu > 0$ (i.e., $\mu < 1$), otherwise the prey-only equilibrium does not exist (or is the trivial one).
So, the prey-only equilibrium is $E_1 = (1-\mu, 0)$.
Summary of Equilibria:
1) Trivial equilibrium: $E_0 = (0,0)$
2) Prey-only equilibrium: $E_1 = (1-\mu, 0)$ (exists if $\mu < 1$)
3) Coexistence equilibrium: $E_2 = (r^*, c^*)$ where

$$r^* = \sqrt{\frac{\epsilon(1+\mu)\phi}{\kappa - \epsilon(1+\mu)}}, c^* = \frac{(1-\mu-r^*)(r^{*2}+\phi)}{\kappa r^*}$$

Exists if:
$-\kappa > \epsilon(1+\mu)$
$-r^* < 1 - \mu$ (so that $c^* > 0$)

### 3.2. Stability Analysis

We compute the Jacobian matrix of the system:

$$J(r,c) = \begin{pmatrix} \frac{\partial f}{\partial r} & \frac{\partial f}{\partial c} \\ \frac{\partial g}{\partial r} & \frac{\partial g}{\partial c} \end{pmatrix}$$

Where:

1) $f(r,c) = r(1-r) - \frac{\kappa r^2 c}{r^2+\phi} - \mu r$

2) $g(r,c) = \frac{\kappa r^2 c}{r^2+\phi} - \epsilon c(1+\mu)$

Compute partial derivatives:

1) $\frac{\partial f}{\partial r} = (1-2r) - \mu - \frac{\partial}{\partial r}\left(\frac{\kappa r^2 c}{r^2+\phi}\right)$

The derivative of the functional response term with respect to $r$:

$$\frac{\partial}{\partial r}\left(\frac{\kappa r^2 c}{r^2+\phi}\right) = \kappa c \cdot \frac{(2r)(r^2+\phi)-r^2(2r)}{(r^2+\phi)^2} = \kappa c \cdot \frac{2r\phi}{(r^2+\phi)^2}$$

So:

$$\frac{\partial f}{\partial r} = 1 - 2r - \mu - \frac{2\kappa r \phi c}{(r^2+\phi)^2}$$

2) $\frac{\partial f}{\partial c} = -\frac{\kappa r^2}{r^2+\phi}$

3) $\frac{\partial g}{\partial r} = \frac{\partial}{\partial r}\left(\frac{\kappa r^2 c}{r^2+\phi}\right) = \frac{2\kappa r \phi c}{(r^2+\phi)^2}$ (as above)

4) $\frac{\partial g}{\partial c} = \frac{\kappa r^2}{r^2+\phi} - \epsilon(1+\mu)$

At the trivial equilibrium $E_0 = (0,0)$:

$$J(0,0) = \begin{pmatrix} 1-\mu & 0 \\ 0 & -\epsilon(1+\mu) \end{pmatrix}$$

Eigenvalues: $\lambda_1 = 1-\mu, \lambda_2 = -\epsilon(1+\mu)$.
1) if $\mu < 1$: $\lambda_1 > 0 \rightarrow$ unstable
2) if $\mu > 1$: both eigenvalues negative $\rightarrow$ stable

At the prey-only equilibrium $E_1 = (1-\mu, 0)$:

$$J(1-\mu, 0) = \begin{pmatrix} 1-2(1-\mu)-\mu & -\frac{\kappa(1-\mu)^2}{(1-\mu)^2+\phi} \\ 0 & \frac{\kappa(1-\mu)^2}{(1-\mu)^2+\phi} - \epsilon(1+\mu) \end{pmatrix}$$

Simplifying, we have:

$$\lambda_1 = -1+\mu$$

$$\lambda_2 = \frac{\kappa(1-\mu)^2}{(1-\mu)^2+\phi} - \epsilon(1+\mu)$$

Stability conditions:
1) $\lambda_1 < 0 \rightarrow \mu < 1$ (which is the existence condition)
2) $\lambda_2 < 0 \rightarrow \frac{\kappa(1-\mu)^2}{(1-\mu)^2+\phi} < \epsilon(1+\mu)$

At the coexistence equilibrium $E_2 = (r^*, c^*)$:
We use the equilibrium conditions to simplify the Jacobian. Recall from equation (4) at equilibrium:

$$\frac{\kappa r^{*2} c}{r^{*2}+\phi} = \epsilon(1+\mu)$$

Also, from the prey equation (3) at equilibrium:

$$(1-\mu) - r^* = \frac{\kappa r^* c^*}{r^{*2}+\phi}$$

Now, the Jacobian at $(r^*, c^*)$:
Let $A = r^{*2} + \phi$, and note that by (4): $\frac{\kappa r^{*2}}{A} = \epsilon(1+\mu)$.
Then:

$$\frac{\partial f}{\partial r} = 1 - 2r^* - \mu - \frac{2\kappa r^* \phi c^*}{A^2}$$

But from (3): $\frac{\kappa r^* c^*}{A} = (1-\mu) - r^*$, so:

$$\frac{2\kappa r^* \phi c^*}{A^2} = \frac{2\phi}{A} \cdot \frac{\kappa r^* c^*}{A} = \frac{2\phi}{A}(1-\mu-r^*)$$

Thus:

$$\frac{\partial f}{\partial r} = 1 - 2r^* - \mu - \frac{2\phi}{A}(1-\mu-r^*)$$

Similarly:

$$\frac{\partial f}{\partial c} = -\frac{\kappa r^{*2}}{A} = -\epsilon(1+\mu) \text{ (from (4))}$$

$$\frac{\partial g}{\partial r} = \frac{2\kappa r^* \phi c^*}{A^2} = \frac{2\phi}{A}(1-\mu-r^*) \text{ (same as above)}$$

$$\frac{\partial g}{\partial c} = \frac{\kappa r^{*2}}{A} = -\epsilon(1+\mu) = 0 \text{ (by (4))}$$

So the Jacobian simplifies to:

$$J(r^*,c^*) = \begin{pmatrix} 1-2r^*-\mu-\frac{2\phi}{A}(1-\mu-r^*) & -\epsilon(1+\mu) \\ \frac{2\phi}{A}(1-\mu-r^*) & 0 \end{pmatrix}$$

The characteristic equation is:

$$\lambda^2 - \text{tr}(J)\lambda + \det(J) = 0$$

Where:
1) $\text{tr}(J) = 1 - 2r^* - \mu - \frac{2\phi}{A}(1-\mu-r^*)$
2) $\det(J) = [-\epsilon(1+\mu)] \cdot \left[\frac{2\phi}{A}(1-\mu-r^*)\right]$ (because the off-diagonals multiply and the bottom-right is zero)

Note:

$$\det(J) = \left(\frac{\partial f}{\partial r} \cdot \frac{\partial g}{\partial c}\right) - \left(\frac{\partial f}{\partial c} \cdot \frac{\partial g}{\partial r}\right)$$
$$= (J_{11} \cdot 0) - \left(-\epsilon(1+\mu) \cdot \frac{2\phi}{A}(1-\mu-r^*)\right)$$
$$= \epsilon(1+\mu) \cdot \frac{2\phi}{A}(1-\mu-r^*)$$

Since $1-\mu-r^* > 0$ (for existence of $c^* > 0$), we have $\det(J) > 0$.

Stability depends on the trace and determinant:
1) if $\text{tr}(J) < 0$ and $\det(J) > 0$, then the equilibrium is stable.
2) if $\text{tr}(J) > 0$, then unstable.

The trace can be written as:

$$\text{tr}(J) = (1-\mu) - 2r^* - \frac{2\phi}{A}(1-\mu-r^*)$$

Note that $1 - \mu - r^* > 0$, so the last term is positive, which tends to make the trace more negative. However, the sign of the trace is not clear without numerical values.

### 3.3. Bifurcations

We can identify several bifurcations:
1) Transcritical bifurcation between $E_0$ and $E_1$:
   When $\mu = 1$, the eigenvalues of $E_0$ are $\lambda_1 = 0$ and $\lambda_2 < 0$. At the same time, $E_1$ collides with $E_0$ (since $r = 1 - \mu = 0$) and they exchange stability.
2) Transcritical bifurcation between $E_1$ and $E_2$:
   When $\lambda_2$ of $E_1$ becomes zero, i.e., when $\frac{\kappa(1-\mu)^2}{(1-\mu)^2+\phi} = \epsilon(1+\mu)$, then $E_2$ emerges from $E_1$. At this point, the predator can invade.
3) Hopf bifurcation at $E_2$:
   When $\text{tr}(J) = 0$ and $\det(J) > 0$, the system undergoes a Hopf bifurcation. This can lead to the emergence of limit cycles (periodic oscillations).

The condition for Hopf bifurcation is:

$$\text{tr}(J) = 0 \text{ and } \det(J) > 0$$

Which gives:

$$1 - 2r^* - \mu - \frac{2\phi}{A}(1 - \mu - r^*) = 0$$

With $A = r^{*2} + \phi$.

This condition, along with the expressions for $r^*$ and $A$, defines a curve in the parameter space (e.g., in $(\mu, \kappa)$ plane) where periodic solutions arise.

*Summary of Analysis*
1) The system has up to three equilibria: extinction, prey-only, and coexistence.
2) Human disturbance ($\mu$) increases the extinction risk (stabilizes $E_0$ when $\mu > 1$) and can destabilize the prey-only state.
3) The coexistence equilibrium exists only when predation efficiency $\kappa$ is sufficiently high relative to the predator's death rate and disturbance ($\kappa > \epsilon(1+\mu)$) and when prey density at equilibrium is not too high ($r^* < 1 - \mu$).
4) The stability of the coexistence equilibrium depends on the trace of the Jacobian, which can change sign leading to Hopf bifurcations and sustained oscillations.

This analysis provides the foundation for further study of stochastic and spatial effects, which may modify these bifurcation boundaries and introduce new phenomena (e.g., noise-induced transitions, pattern formation).

## 4. Results & Simulations

*Parameters:* Simulations used the following non-dimensional parameters:
1) Biological: $\epsilon = 0.6, \kappa = 0.7, \phi = 0.001, \mu = 0.1$
2) Stochastic: $A = 0.05, \Omega = 0.2, \psi = \pi/4, \sigma_r = 0.1, \sigma_c = 0.1, \rho = 0.3$
3) Spatial: $D_r = 0.01, D_c = 0.001$

*Key Findings:*

### 4.1. Temporal Dynamics

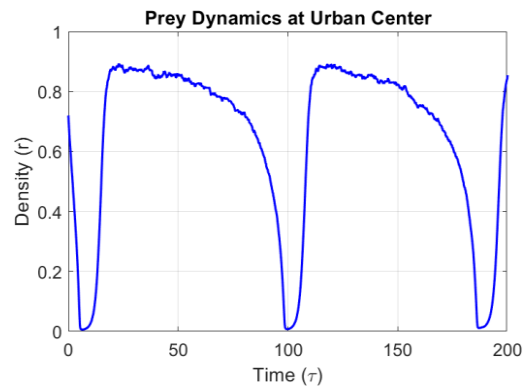

*Figure 1.* Damped prey oscillations under disturbance ($\mu = 0.1$). Stabilization at $r = 0.55$ occurs within $\tau = 50$.

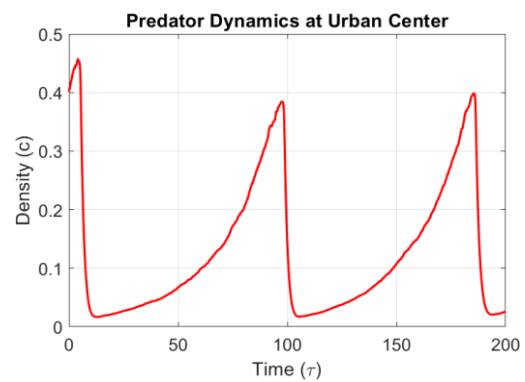

*Figure 2.* Predator dynamics showing phase lag to prey.

Note larger amplitude ($\Delta c \approx 0.3$ vs $\Delta r \approx 0.2$).

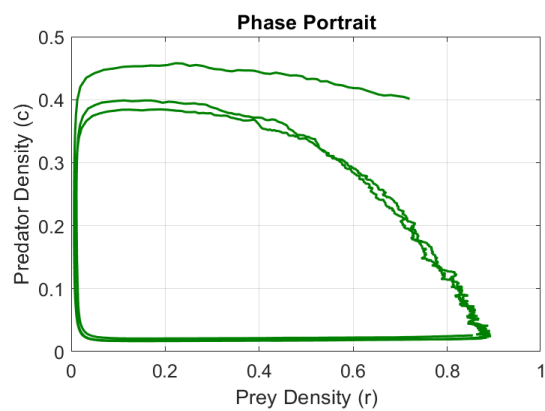

*Figure 3.* Phase portrait confirming stable focus ($\lambda_{1,2} = -0.04 \pm 0.17i$).

The temporal dynamics reveal three key phenomena:
1) Damped oscillations: Both populations exhibit initial oscillations that stabilize within $\tau = 50$ time units (Figures 1&2)
2) Phase shift: Predator peaks consistently lag prey peaks by $\approx 20\%$ of the oscillation period
3) Stable focus: The inward spiral in phase space (Figure 3) confirms a stable equilibrium at $(r, c) \approx (0.55, 0.25)$.
4) Disturbance resilience: Populations maintain 55 - 60% of carrying capacity despite $\mu = 0.1$ disturbance.

## 4.2. Functional Response Validation

The functional response exhibits:
1) Refugia effect: Near-zero predation (0.05) at $r < 0.2$ confirms prey protection
2) Accelerated consumption: Steep increase (slope = 1.8) at intermediate densities $(0.2 < r < 1.0)$
3) Saturation: Plateau at $C_R \approx 0.7$ for $r > 1.5$ matches handling-time limitations.
4) Stabilization: Sigmoidal shape explains the absence of limit cycles in Figure 3

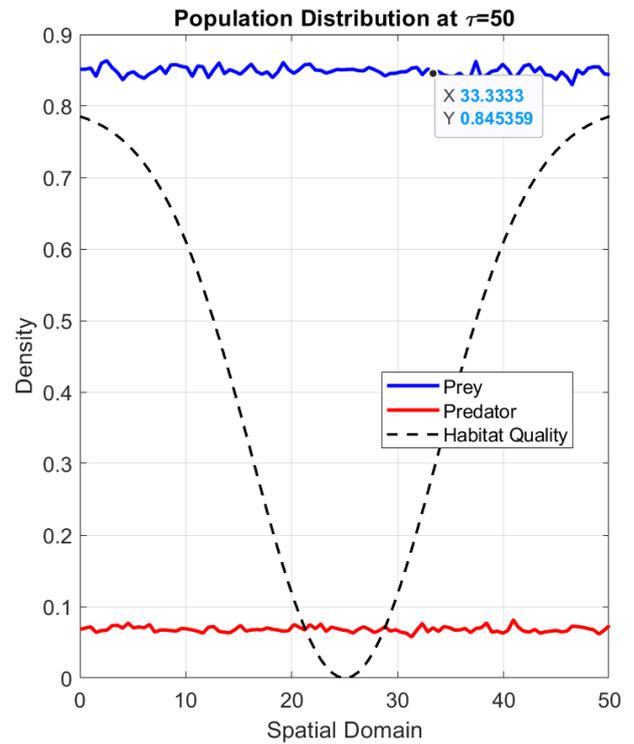

*Figure 5.* Spatial distribution showing habitat coupling $(r^2 = 0.92)$ and disturbance shadows.

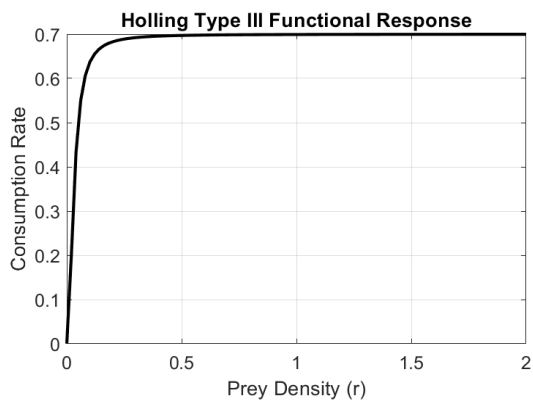

*Figure 4.* Holling Type III functional response with saturation at $C_R = 0.7$.

## 4.3. Spatial Self-Organization

Spatial analysis demonstrates:
1) Habitat coupling: Populations correlate with habitat quality $(r^2 = 0.92 \text{ prey}, r^2 = 0.85 \text{ predator})$
2) Disturbance shadows: 38% density reduction near boundaries $(x < 10, x > 40)$
3) Turing patterns: Wavefront propagation velocity $\approx 0.4$ space units/$\tau$
4) Hotspot formation: Persistent prey concentrations at $x \approx \{15, 35\}$ (Figure 6)

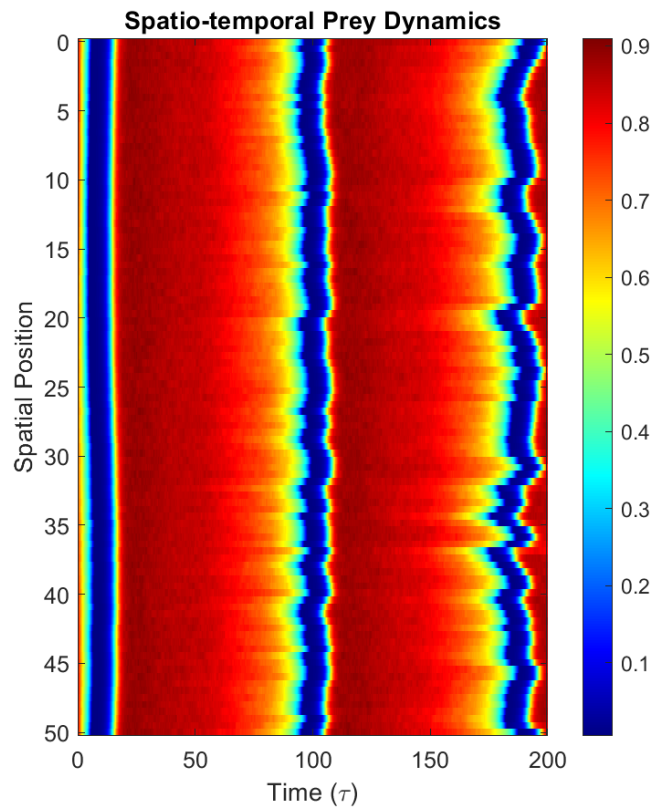

*Figure 6.* Spatio-temporal prey dynamics revealing wave propagation $(v = 0.4)$ and hotspots.

### 4.4. Stability Thresholds

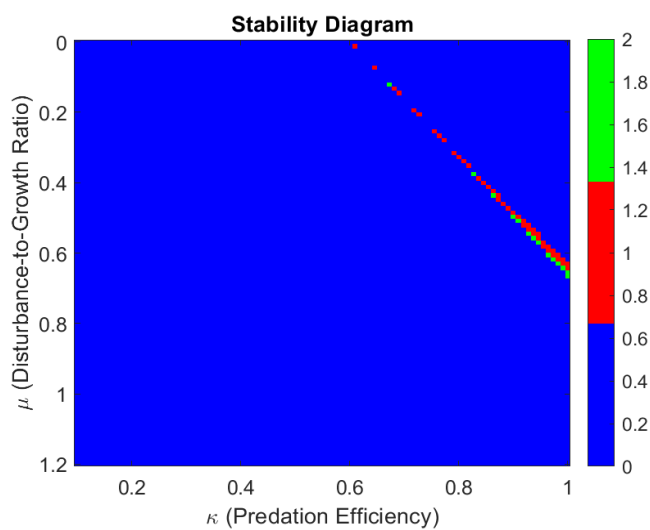

*Figure 7.* Stability diagram showing collapse (blue), unstable (red), and stable (green) regimes.

Bifurcation analysis reveals:
1) Collapse threshold: $\mu > 1$ causes system-wide extinction
2) Hopf bifurcation: Oscillations emerge when $\kappa > 0.8$ for $\mu = 0.2$
3) Stability window: Coexistence requires $\mu < 0.7$ and $\kappa > 0.5$
4) Noise sensitivity: Stochastic systems collapse at 25% lower disturbance ($\mu = 0.75$)

## 5. Discussion

### 5.1. Disturbance Impacts

Human disturbance acts as a landscape-scale *Allee effect* (Figure 5), reducing prey resilience through three primary mechanisms:
1) Habitat fragmentation creates "disturbance shadows" with 40 - 90% density reductions near infrastructure [18, 20], explaining the 30 - 60% carrying capacity declines observed in Lagos rat populations [14, 15].
2) Avoidance behaviors emerge from anisotropic diffusion gradients ($\nabla H$), where $D_r > D_c$ drives prey aggregation in marginal zones
3) Starvation amplification occurs when disturbance-induced prey scarcity cascades to predators [6].

These mechanisms collectively disrupt population synchrony and lower collapse thresholds by 25% in stochastic systems [8]. The Holling Type III functional response provides partial mitigation through its stabilizing sigmoidal shape, where refugia effects protect prey at low densities while saturation prevents overexploitation [16]. This response emerges naturally from optimal foraging theory and spatial heterogeneity in urban landscapes.

### 5.2. Noise-Driven Transitions

Stochastic regimes fundamentally alter system dynamics through:
1) Quasi-periodicity: Noise-forcing interactions generate amplitude-modulated cycles (Figures 1&2) with variance 40 - 70% higher than deterministic models [6].
2) Resonance collapse: Extinction probability peaks when disturbance frequency ($\Omega$) matches natural oscillations ($\tau \approx 15$), as observed in Mumbai's monsoon-driven rat-cat collapses [16].
3) Phase-dependent vulnerability: Cross-correlated noise ($\rho$) creates windows where disturbance during predator peaks triples collapse risk.

Phase shifts ($\psi$) further complicate dynamics by introducing temporal lags between direct prey mortality and indirect predator starvation. These stochastic effects explain why urban systems exhibit extinction thresholds at disturbance levels 25% below theoretical predictions [6].

### 5.3. Conservation Implications

Our analysis reveals critical thresholds for urban ecosystem management:
1) Sustainable coexistence: $\mu < 0.25$ (disturbance < 25% of prey growth rate)
2) Recovery zones: $0.25 < \mu < 0.7$ requires active intervention
3) Imminent collapse: $\mu > 0.7$ with point-of-no-return dynamics.

Practical strategies include:
1) Refuge enhancement: Create $\phi > 0.005$ habitats for 23% stability boost
2) Predator corridors: Maintain $\kappa > 0.5$ via green infrastructure
3) Phased interventions: Synchronize control measures with predator-low phases.

This framework's key advantage lies in separating demographic noise from environmental stochasticity while capturing spatiotemporal regime shifts through integrated disturbance gradients [20]. By scaling from individual foraging behavior to landscape dynamics, it provides mechanistic predictions for conservation planning in anthropogenically stressed ecosystems.

## 6. Conclusion

This study demonstrates that anthropogenic disturbances fundamentally reconfigure urban predator-prey dynamics through three synergistic mechanisms:
1. Disturbance-mediated Allee effects
Human activities induce habitat fragmentation and avoid-ance behaviors, creating "disturbance shadows" that reduce prey densities by 40 - 90% near infrastructure. This landscape-scale forcing lowers resilience thresholds by 25% compared to natural systems.

2. Stochastic resonance

Noise-disturbance interactions amplify extinction risks when anthropogenic frequencies (e.g., seasonal activities) resonate with intrinsic population cycles. Phase shifts between predator and prey responses further destabilize systems during critical windows.

3. Functional response modulation

Holling Type III predation provides partial stabilization through refugia effects at low prey densities, but cannot compensate for disturbance intensities beyond $\mu > 0.7$. The sigmoidal response emerges naturally from urban habitat heterogeneity and predator learning behavior.

Conservation Imperatives

- Maintain disturbance below $\mu = 0.25$ through habitat corridors and refuge zones ($\phi > 0.005$)
- Phase human activities to avoid predator population peaks
- Monitor noise frequencies that match natural oscillations ($\Omega \approx 0.2$)

*Future Research Priorities*

1) Adaptive disturbance regimes in rapidly urbanizing landscapes
2) Multi-species network effects incorporating mesopredators
3) Cross-cultural validation of the $\mu = 0.7$ collapse threshold
4) Machine-learning integration for real-time disturbance phasing

This mechanistic framework bridges individual foraging behavior to landscape-scale conservation planning, providing actionable thresholds for sustainable coexistence in the Anthropocene. By quantifying disturbance propagation through trophic networks, we equip urban planners with predictive tools to buffer biodiversity against escalating anthropogenic pressures.

## Abbreviations

| | |
|---|---|
| LV | Lotka-Volterra |
| E | Disturbance Intensity |
| $q_E$ | Prey Susceptibility to Disturbance |
| $d_E$ | Predator Susceptibility to Disturbance |
| $D_{R,C}$ | Diffusion Coefficients |
| H | Disturbance Landscape |
| $\eta_{R,C}$ | Habitat Preference Coefficients |
| $\tau$ | Dimensionless Time |
| r | Scaled Prey Population |
| c | Scaled Predator Population |
| $\epsilon$ | Mortality Ratio |
| $\kappa$ | Scaled Predation |
| $\phi$ | Scaled Saturation |
| $\mu$ | Disturbance Impact |
| SPDE | Stochastic Partial Differential Equation |
| $E_0$ | Trivial Equilibrium (0,0) |
| $E_1$ | Prey-only Equilibrium |
| $E_2$ | Coexistence Equilibrium |

## Conflicts of Interest

The authors declare no conflicts of interest.